\documentstyle[11pt,newpasp,twoside,epsf]{article}
\begin{document}

\title{Frequency of Gas-Stellar Counterrotation}
 
\author{S. J. Kannappan and D. G. Fabricant}
\affil{Harvard-Smithsonian CfA, 60 Garden St MS-20, Cambridge, MA 02138}

\begin{abstract}
We search for bulk counterrotation of the gas and stars in 67 galaxies
of all types and a wide range of luminosities.  Bulk counterrotation
characterizes $\sim$25--30\% of E/S0's with extended gas, but at most
a few percent of Sa--Sbc spirals.  For S0's, the frequency of
counterrotation we derive agrees with previous work, but we sample
significantly fainter luminosities.  Thus the agreement suggests that
similar formation mechanisms may operate over a wide range of physical
scales.
\end{abstract}

\section{Sample}
The survey sample (Figure~1) consists of 67 galaxies drawn from the
Nearby Field Galaxy Survey (NFGS), a survey of $\sim$200 galaxies
including all morphological types in their natural abundance and
spanning luminosities from M$_{\rm B}\sim$ $-$23 to $-$15 (Jansen et al.\
2000).  Our sample includes all of the NFGS E/S0's with extended gas
emission, a representative majority of the Sa--Sbc spirals, and a few
later types.  Three S0's show minimal rotation with uncertain sense.

\section{Results}
Counterrotators are circled in Figure~1.  Four are E/S0's, totalling
24(29)\% of all sample E/S0's with(without) the 3 uncertain S0's.  One
is an Im, for which the confidence of our counterrotation claim is
$\sim$2.2$\sigma$.  Full details appear in Kannappan \& Fabricant
(2000).  We stress that our detections represent a lower limit,
because we cannot separate multiple kinematic components.

\begin{figure}
\plottwo{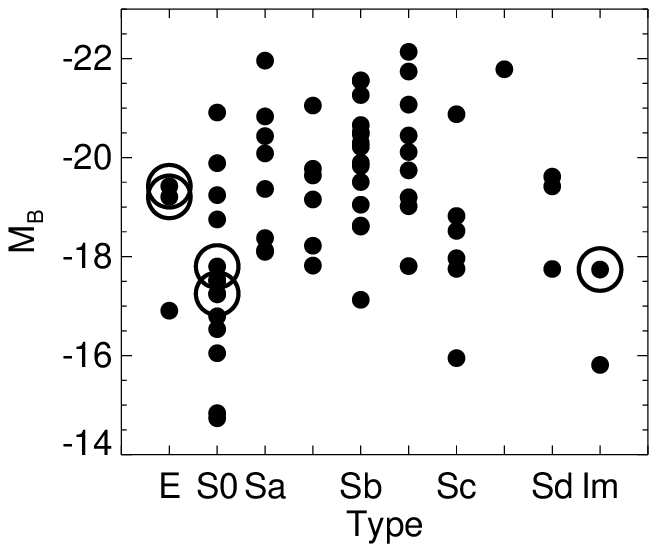}{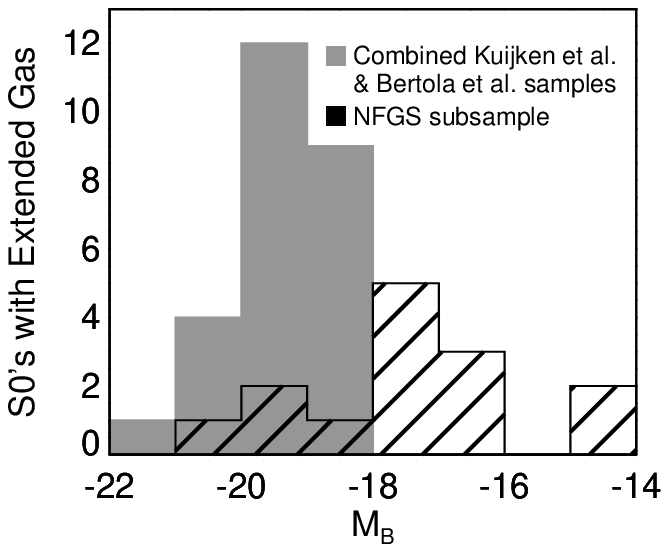}
\caption{Left: The search sample, with counterrotators circled.
Right: Luminosity distributions for the 14 S0's in our sample and for
the Bertola et al.\ (1992) and Kuijken et al.\ (1996) studies.}
\end{figure}

\section{Discussion}
Bertola, Buson, \& Zeilinger (1992) and Kuijken, Fisher, \& Merrifield
(1996), find gas-stellar counterrotation frequencies of $\sim$20--25\%
for emission line S0's.  We find 14(18)\% with(without) the 3
uncertain S0's.  This agreement, despite a strong difference in sample
luminosity distributions (Figure~1), suggests that similar mechanisms
form gas rich S0's over a wide range of physical scales.

Among 38 Sa--Sbc spiral galaxies, we find no clear cases of bulk
counterrotation, implying a counterrotation frequency of $<$8\% (95\%
confidence).  However, counterrotation cannot be ruled out in the
inner parts of one Sa (NGC~4795).  This galaxy shows clear evidence of
gas-stellar decoupling at large radii (Figure~2), possibly related to
satellite accretion.  Jore (1997) finds a bulk counterrotation
frequency of $\sim$17\% in a sample of 23 bright Sa's, consistent with
our results given our Sa sample size of 8 (see also Haynes et al.\
2000).

\begin{figure}
\plotone{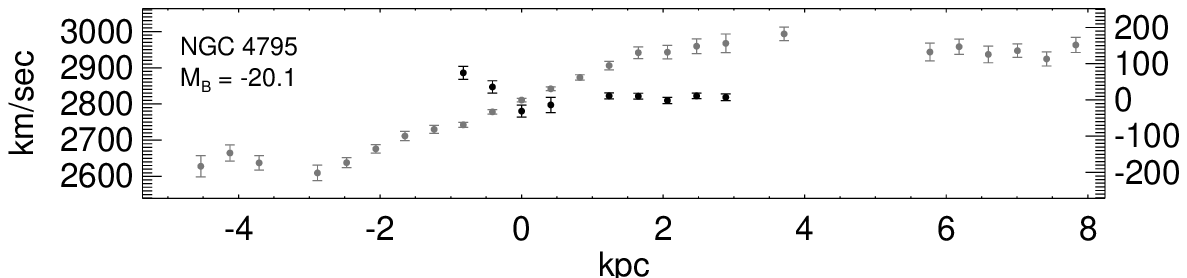}
\caption{Gas (black) and stellar (grey) kinematics for NGC~4795.}
\end{figure}

As argued in Kannappan \& Fabricant (2000), mergers and accretion
provide the simplest explanation for the counterrotators' properties,
particularly their early type morphologies, significant HI masses, and
lack of obvious close neighbors.  The signatures of minor retrograde
accretion events may be washed out in gas rich spirals, while
more substantial mergers are likely to produce E/S0/Sa galaxies.
Dwarf-dwarf mergers may help to explain the faint S0 population.

\acknowledgements We thank R.~Jansen for sharing data and expertise.
S.~J.~K. acknowledges support from a NASA GSRP Fellowship.


\begin{references}

\reference {Bertola}, F., {Buson}, L.~M., \& {Zeilinger}, W.~W. 1992, \apjl, 401, L79

\reference {Haynes}, M.~P., {Jore}, K.~P., {Barrett}, E.~A., {Broeils}, A.~H., \&
  {Murray}, B.~M. 2000, \aj, 120, 703

\reference {Jansen}, R.~A., {Franx}, M., {Fabricant}, D., \& {Caldwell}, N.
  2000, \apjs, 126, 271

\reference {Jore}, K.~P. 1997, PhD thesis, Cornell University

\reference {Kannappan}, S.~J. \& {Fabricant}, D.~G.  2000, to appear in \aj

\reference {Kuijken}, K., {Fisher}, D., \& {Merrifield}, M.~R. 1996, \mnras, 283, 543

\end{references}
\end{document}